# 一种改进的点在凸多边形内外判断算法 *

孙翊轩[1]，朱哲皓[2]

(1.东营市第一中学, 山东 东营 257100;　2.上海工程技术大学, 上海 200093)

**摘　要**：为判断点在简单凸多边形内外，提出一种基于垂线段相交判断，令点确定在两条直线之间，将判断点在简单凸多边形内外的问题转化为判断点在四边形内外的问题，再通过射线法判断判断点在四边形内外，其期望复杂度小于等于 O(N)。实验结果表明，与传统直接判断的方法比较，该算法线段求交次数更少，有效提高判断效率。

**关键词**：任意简单多边形；计算几何；几何检索；点与多边形关系

**中图分类号**：TP391.72

## An improved judgement algorithm of point in-out convex polygons

Sun Yixuan[1], Zhu Zhehao[2]

(1. *DongYing No.1 Middle School, Dongying* Shandong 257100, *China*; 2. *Shanghai University Of Engineering Science, Shanghai* 200093, *China*)

**Abstract**:This essay introduced a method to judge whether a point is inside or outside a simple convex polygon by the intersection of a vertical line. It determines a point to an area enclosed by two straight lines and then converts the problem into deciding whether a point is inside or outside a quadrilateral. After that, use the ray method to verify it. The expected complexity of this algorithm is equal to or less than O(N). As the experimental results, the algorithm has fewer intersections and significantly improves the verification efficiency.

**Key words**: arbitrary simple polygon; computational geometry; geometry search; point and polygon relationship

## 0　引言

判断点在多边形内外是计算几何重要的基本内容，也是在地理测绘、游戏引擎等方面有广泛应用的课题。点在多边形内外的判断即判断指定一点是否在指定多边形内或在指定多边形外。

目前已被提出的方法可以分为两类：

a) 第一类为不经过任何预处理操作，如射线法[1~4]、三角剖分法[3~4]、夹角之和检测法[5~6]等。上述方法均为运用多边形的性质，不需要多余的内存占用，若需判断的多边形边数为N，其最坏复杂度均为O(N)。

b) 第二类为经过预处理建立索引，如网格索引法[7~8]、梯形化[9~11]扫描线二分BSP搜索法[12]等，都是通过建立索引，降低每一次查询的复杂度。虽然这一类算法之间的时间复杂度存在差别，但最坏复杂度均为O(N)。

本文提出一种不进行预处理、针对凸多边形的算法，通过作垂线段确定点的初步位置，再利用凸多边形特点将点锁定在更小区间中，以提高在简单凸多边形中的判断效率。

## 1　前置定理

本文提出的算法需要证明以下推论：

如图 1，对于两条交于 F 的不共线的直线，过其中一条直线上一点 M 作这条直线的垂线段交另一条直线于一点 G，这两点确定的线段 GM 必定在这两条直线的 GM 一侧所围出的区间（即$l_1$下侧区间与$l_2$上侧区间的交集）内。

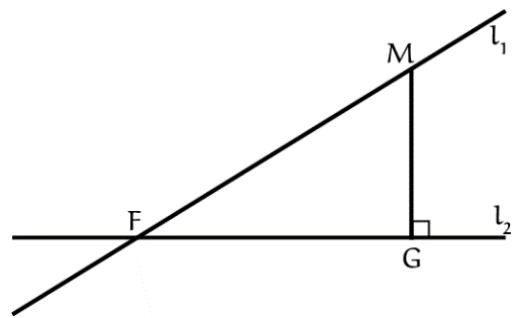

图 1　GM 与直线示意图

Fig.1　GM and straight-line schematic diagram

---

## 1.1 定理证明

如图 1，在一个平面直角坐标系中有两条任意不重合的直线：$l_1$和$l_2$，过点 M 作$l_2$的垂线段 MG 交$l_2$于点 G。

$l_1$的点方向式为：

$$\frac{x-x_g}{u_1} = \frac{y-y_g}{v_1} \quad (1)$$

（$x_g$和$y_g$是$l_1$上的点 G 的横坐标和纵坐标）

$l_2$的点方向式为：

$$\frac{x-x_m}{u_2} = \frac{y-y_m}{v_2} \quad (2)$$

（$x_m$和$y_m$是$l_2$上的点 M 的横坐标和纵坐标）

当$l_1$的斜率≥$l_2$且$l_1$、$l_2$的斜率均≥0 时，将$l_1$和$l_2$的交点记作 F，易得 F 的横坐标为：

$$x_f = \frac{y_g u_1 u_2 - y_m u_1 u_2 + x_g v_1 u_2 - x_m v_2 u_1}{v_2 u_1 - v_1 u_2} \quad (3)$$

首先讨论$x_g \geq x_f$时的情况，直线 GM 可以表示为：

$$l_{GM}: \frac{x-x_g}{x_g-x_m} = \frac{y-y_g}{y_g-y_m} \quad \text{或} \quad l_{GM}: \frac{x-x_m}{x_g-x_m} = \frac{y-y_m}{y_g-y_m} \quad (4)$$

此时线段 GM 可以表示为：

$$\frac{x-x_g}{x_g-x_m} = \frac{y-y_g}{y_g-y_m} \quad (x_m \geq x \geq x_g) \quad (5)$$

也可以表示为：

$$\frac{x-x_m}{x_g-x_m} = \frac{y-y_m}{y_g-y_m} \quad (x_m \geq x \geq x_g) \quad (6)$$

将(5)、(6)式稍加变形可得：

$$y = \frac{(x-x_g)(y_g-y_m)}{x_g-x_m} + y_g \quad (x_m \geq x \geq x_g) \quad (7)$$

$$y = \frac{(x-x_m)(y_g-y_m)}{x_g-x_m} + y_m \quad (x_m \geq x \geq x_g) \quad (8)$$

$l_1$和$l_2$在$x_g \geq x_f$上围出的区间可以表示为：

$$\begin{cases} y \leq \frac{(x-x_g)v_1}{u_1} + y_g \\ y \geq \frac{(x-x_m)v_2}{u_2} + y_m \\ x \geq \frac{y_g u_1 u_2 - y_m u_1 u_2 + x_g v_1 u_2 - x_m v_2 u_1}{v_2 u_1 - v_1 u_2} \end{cases} \quad (9)$$

若要证明线段 GM 完全包含于$l_1$和$l_2$在$x_g \geq x_f$上围出的区间，只需证明(7)式或(8)式包含于(9)式，即：

$$\frac{(x-x_m)v_2}{u_2} + y_m \leq \frac{(x-x_g)(y_g-y_m)}{x_g-x_m} + y_g \leq \frac{(x-x_g)v_1}{u_1} + y_g \quad (x_m \geq x \geq x_g) \quad (10)$$

将(10)式中不等式拆分成两部分分别作差：

$$\frac{(x-x_g)v_1}{u_1} + y_g - \frac{(x-x_g)(y_g-y_m)}{x_g-x_m} - y_g \quad (11\text{-}1)$$

$$\frac{(x-x_g)(y_g-y_m)}{x_g-x_m} + y_g - \frac{(x-x_g)v_2}{u_2} - y_m \quad (11\text{-}2)$$

化简(11-1)式、(11-2)式并使用(8)式等价代换与(7)式相同的部分，可得：

$$\frac{(x-x_g)v_1}{u_1} - \frac{(x-x_g)(y_g-y_m)}{x_g-x_m} \quad (12\text{-}1)$$

$$\frac{(x-x_m)(y_g-y_m)}{x_g-x_m} - \frac{(x-x_g)v_2}{u_2} \quad (12\text{-}2)$$

由$l_1$的斜率≥$l_2$且$l_1$、$l_2$的斜率均≥0 可知$u_1$、$v_1$同号，$u_2$、$v_2$同号，直线 GM 的斜率<0。则由$x_m \geq x \geq x_g$又可得：

$$\begin{cases} x - x_m \leq 0 \\ x_g - x_m \leq 0 \\ x - x_g \geq 0 \end{cases} \quad (13)$$

又由直线 GM 的斜率<0 得到$y_g - y_m \geq 0$条件成立，则下式成立：

$$\begin{cases} \frac{(x-x_g)(y_g-y_m)}{x_g-x_m} + y_g \leq \frac{(x-x_g)v_1}{u_1} + y_g \\ \frac{(x-x_m)v_2}{u_2} + y_m \leq \frac{(x-x_g)(y_g-y_m)}{x_g-x_m} \end{cases} \quad (14)$$

综上，(10)式成立，即线段 GM 完全包含于$l_1$和$l_2$在$x_g \geq x_f$上围出的区间。当$x_g \leq x_f$时证明步骤同上，则可得在当前情况下，线段 GM 完全包含于$l_1$和$l_2$在$x_g \leq x_f$上围出的区间。

## 1.2 定理推广

当$l_1$、$l_2$的斜率为任意情况时，可以对其不同可能性按照上述证明方式推广并使用前文 1.1 的方法证明：

a) $l_1$的斜率≥$l_2$的斜率且$l_1$、$l_2$的斜率均≥0
b) $l_1$的斜率≥$l_2$的斜率且$l_1$、$l_2$的斜率均<0
c) $l_1$的斜率≥0 且$l_2$的斜率<0
d) $l_1$的斜率≤$l_2$的斜率且$l_1$、$l_2$的斜率均≥0
e) $l_1$的斜率≤$l_2$的斜率且$l_1$、$l_2$的斜率均<0
f) $l_1$的斜率<0 且$l_2$的斜率≥0

综上可得，线段 GM 完全包含于任意$l_1$和$l_2$围出的区间，即 1.1 中的推论成立。

# 2 一种改进的点在凸多边形内外的判断算法

## 2.1 定理拓展

在图 2 中，有一简单凸多边形和一点 P，一点向随机一条边作垂线段 PG，由前文 1 中得到的结论，可得到图中的垂线段的垂足 G 与被垂直线段的两条邻边外侧两点连线交点 M 一定在被垂直线段所在直线与被垂直线段的两条邻边外侧两点所在直线所围出的区间之中（如图 2 灰色区域）。

若该垂线段与被垂直线段的两条邻边另外两点连线相交时（如图 2），点 P 一定不在线段 GM 上，反之，若该垂线与两条邻边外侧两点连线没有交点（如图 3），则点 P 一定在线段 GM 上，又因为 GM 一定在被垂直线段所在直线与被垂直线段的两条邻边外侧两点所在直线所围出的区间之中，故 P 也在该区间中。

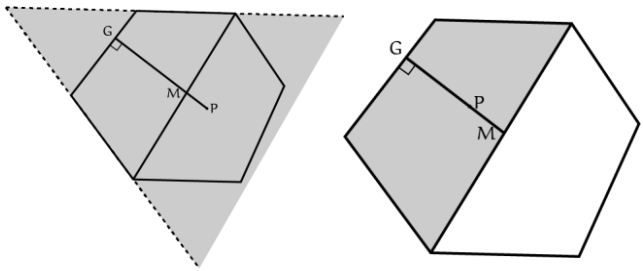

图 2  P 不在 GM 上时　　　图 3  P 在 GM 上时

Fig.2 While Point P is not　　Fig.3 While Point P is

on line segment GM　　　　on line segment GM

因此，如果线段 PG 与被垂直线段邻边外侧两点连线没有交点，则 P 一定在被垂直线段所在直线与被垂直线段两条邻边外侧两点所在直线围出的区间内。此时若需要判断点 P 是否在该凸多边形内，仅需判断 P 是否在被垂直线段两点与邻边外侧两点依次连接构成的四边形中（即图 3 中灰色部分中凸多边形内部分）即可。

### 2.2 算法流程

对于所有凸多边形应用 2.1 中结论，提出一种改进的点在凸多边形内外判断算法：

在每次判断点是否在凸多边形内时，抽取任意一条边由该点作垂线段。若这条垂线段与被垂直线段的邻边另外两点连线无交点，则按照射线法判断该点是否在四边形内，将结果拓展到凸多边形即可，关于射线法的判定方法及优化已在文献[1~4]中详细阐述，在此不多加赘述。

若上述判断结果为有交点，则继续更换线段进行判断，继续进行如上步骤。如图 4，若在最后每条垂线段与对应线段均有交点，此时因为该点完全不处在任何两条线段所构成的区间内并且垂线段与连边均有交点，若点在多边形外部，则垂线段一定不会与多边形内部的全部线段有交点，即该点一定在多边形内部。

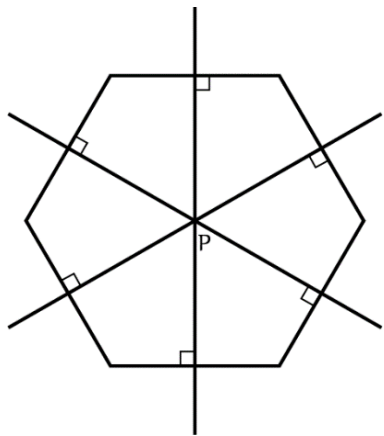

图 4  垂线段与线段均有交点的情况

Fig.4 While all vertical lines and line segments have intersection points

算法流程如图 5 所示：

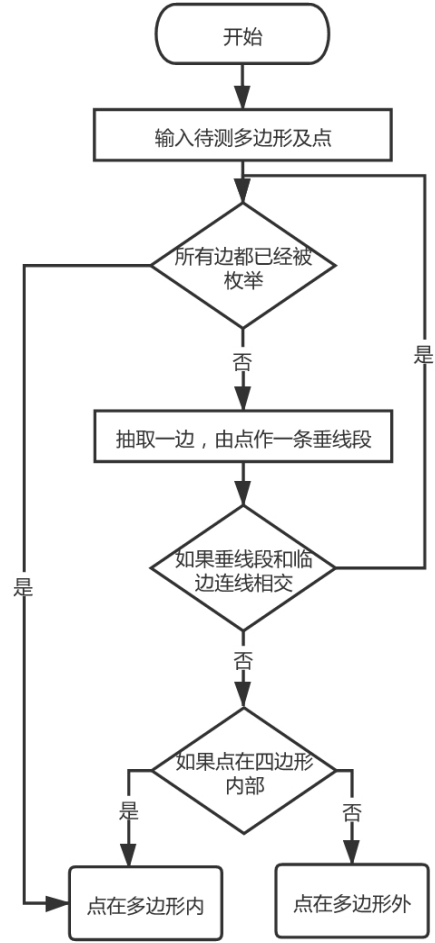

图 5  算法流程图

Fig. 5 Flowchart of algorithm

### 2.3 特殊情况

在进行作垂线段或判断点在多边形内外时会出现一些特殊情况，本节对这些特殊情况进行列举并提出针对特殊情况的解决方法。

a) 点在凸多边形边或点上

如图 6，点 M 若在凸多边形的边或点上时，对该点所在凸多边形内的线段作垂线段时，垂足 G 与 M 本身重合，无法判断垂线段是否与被垂直线段相交。

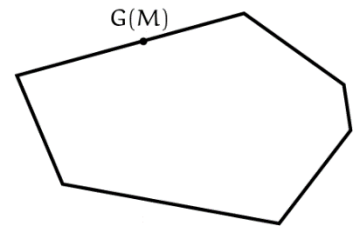

图 6  点在凸多边形上

Fig.6 While the point on the convex polygon

为解决这种问题，可以将该点视为垂线段的两端点，即创造出一条长为 0 的垂线段。在射线法判断点是否在多边形内外时，可先通过判断点是否在该四边形中处在凸多边形内的三条线段上来判断点是否在该凸多边形上。

b) 点与所选取的线段无法作垂线段

如图 7，点 M 与随机抽取的线段 AB 之间，无法作一条合法的垂线段，此时无论点 M 在多边形的任何位置，都不能确定其与多边形的内外位置关系。

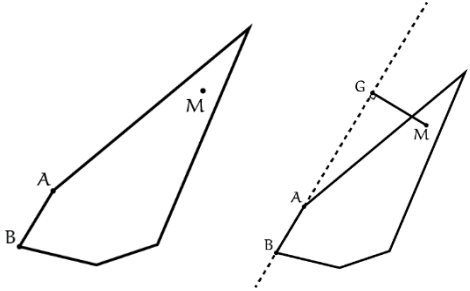

图 7　无法作垂线段的情况　　图 8　向线段所在直线作垂线段

Fig.7 Can not make a vertical line　　Fig.8 Make a vertical line

segment circumstance　　　　　segment to the straight line

对于解决该问题，如图 8，可以将线段 AB 延长作为一条直线，由 M 向直线 AB 作出一条垂线段，再判断该垂线段与被垂直线段邻边外端两点的连线是否相交，即可判断点是否在两直线所围成的区间内。

### 2.4 算法复杂度分析

本文提出的算法可分为两个阶段，即作一条合法垂线段（即该垂线段与被垂直线段的邻边外侧两点连线没有交点）和判断点是否在四边形内两阶段，分析如下：

a) 时间复杂度

考虑作合法垂线段的阶段，程序在找到一条合法垂线段后即进入下一阶段，因此复杂度为找到一条合法垂线段的复杂度。

对于一点 P，设对于一个有 N 条边的多边形 G，P 向多边形所有边作出垂线段中的合法垂线段数为 $\sigma(P,G)$，找到一条合法垂线段所需的期望复杂度为 E。

第一次作垂线段时，若该垂线段为合法垂线段，则进入之后流程的概率为 0 即期望为 0，本次复杂度为 1；若失败，则之后期望为 E，概率为 $1 - \frac{\sigma(P,G)}{N}$，即可得：

$$E = 1 + \frac{N-\sigma(P,G)}{N}E \qquad (15)$$

将(15)式整理可得：

$$E = \frac{N}{\sigma(P,G)} \qquad (16)$$

考虑判断点是否在四边形内的阶段，因为四边形的边数为 4，因此该阶段复杂度为 O(4)[1]，对渐近复杂度无影响。

综上所述，再结合实际情况中合法垂线段数量满足 $\sigma(P,G) \in [0,N] \cap Z$，可以得出本文提出的改进的点在凸多边形内外的判断算法的时间期望复杂度为：

$$O_{G(P,N)} = \frac{N}{\sigma(P,G)} (\sigma(P,G) \in [0,N] \cap Z) \qquad (17)$$

将上文中提及的 N 作为 y 轴，$\sigma(P,G)$ 作为 x 轴，期望复杂度 E 作为 z 轴，根据(17)式作出图 8 中的改进算法复杂度图象：

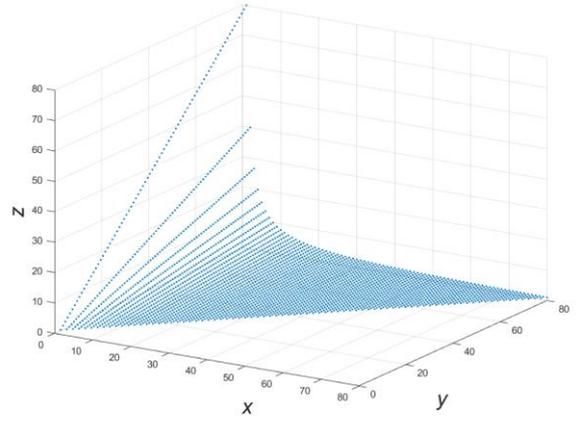

图 8　改进算法复杂度图象

Fig.8 Complexity graph of improved algorithm

同时根据传统方法[1-6]的复杂度 $O_G(P,N) = N$ 作出图 9 中的传统算法复杂度图象：

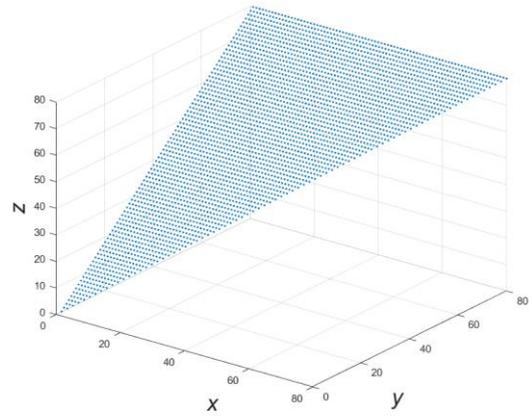

图 9　传统算法复杂度图象

Fig.9 Complexity graph of tranditional algorithm

由图象分析可知,多边形中可以找到的合法垂线段越多，改进算法的期望复杂度越低，且显著优于传统方法，在找不到合法垂线段时，改进算法与传统算法的时间复杂度相同。因此一旦多边形中出现了合法垂线段，使用改进算法判断点在多边形内外效率极高。

b) 空间复杂度

该算法的空间复杂度与多边形的边数无关，算法过程中仅需要定义少量临时变量，因此本文提出的改进的点在凸多边形内外的判断算法的空间复杂度为 O(1)。

## 3　实验结果对比分析

通过 C++实现了该算法，以传统射线法[1]与传统三角剖分法[3]为对照，运行环境为 Linux 5.9 操作系统，CPU 为 Intel Core i5-10th 2.11 GHz，内存 16GB。在一个简单凸多边形的外接矩形中抽取 10000 个点平均分为 10 个点集，依次输入算法程序中，记录算法依次通过上述点集所需的总时间，以传统射线法判断第一个点集的时间作为基准，绘制出如图 10 的点集个数—判断所需时间示意图：

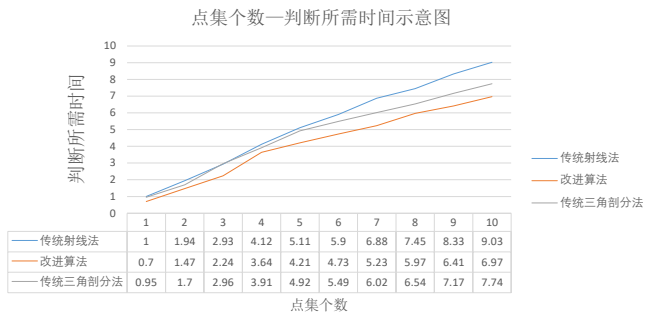

图 10 点集个数—判断所需时间示意图
Fig.10 Point sets-time diagram

再确定一点，随机生成 10 个凸多边形集，第 1~3 集中凸多边形边数为 100 条，第 4~6 集中为 500 条，第 7~9 集中为 1000 条，第 10 集中为 2000 条，依次输入算法程序中，记录算法依次通过上述多边形集的总时间，以传统射线法判断第一个多边形的时间作为基准，绘制出如图 11 的多边形集个数—判断所需相对时间示意图：

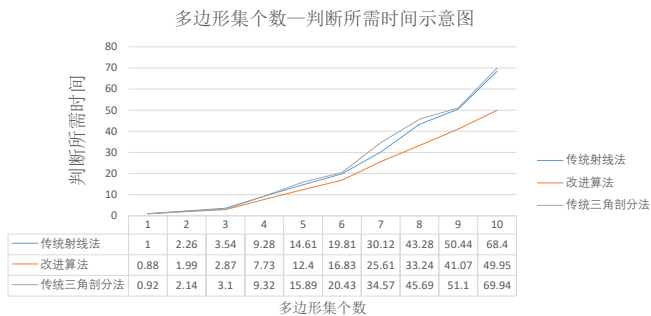

图 11 多边形集个数—判断所需时间示意图
Fig.11 Polygons sets-time diagram

图 11 中三种算法效率差别不大，原因是在边数不断增多时，如果该点在凸多边形靠中间的位置，改进算法找到合法垂线段较困难。若找不到合法垂线段，改进算法与传统方法的时间复杂度将相同，但一旦找到合法垂线段，复杂度一定低于 O(N)，因此实验结果中随着边数逐渐增多，改进算法仍领先于传统方法。

综上实验结果可以得到以下结论：本文提出的改进算法对简单凸多边形的判断效率较好，对于正常凸多边形的判断效率普遍优于传统方法。

## 4 结束语

为解决点在凸多边形内外判断的效率，本文提出了一种改进的点在凸多边形内外判断的算法。该算法先采用作垂线段的方法确定点所在的区间，并将判断点在凸多边形内外简化为判断点在四边形内外。最后，实验结果表明，该算法在点数和多边形边数不断增加的情况下运行效率均优于同类算法。在未来的研究中，一方面，应寻找效率更高的方式替代垂线段相交判断来提升效率，另一方面，应尝试通过预处理的方式让找到合法垂线段的速度更快、准确率更高。